\documentclass[12pt]{iopart}

\usepackage{iopams}  
\begin{document}

\title[A Mass-Shell Model for CBCs]{A Mass-Shell Model of Compact Binary Coalescence}

\author{Noah M. MacKay
\\ ORCID ID: 0000-0001-6625-2321}

\address{Institut für Physik und Astronomie, Universit\"at Potsdam,\\ Karl-Liebknecht-Str. 24/25, 14476 Potsdam, Germany}
\ead{noah.mackay@uni-potsdam.de}
\vspace{10pt}
\begin{indented}
\item[]\today
\end{indented}

\begin{abstract}
The final pulse of gravitational wave (GW) emission is released at the peak of the chirp rise before compact binary merger. LIGO detections since GW150914 reveal a correlation between the radiated energy $E_\mathrm{rad}$ and the ad hoc scaling of one-tenth of the chirp mass $\mathcal{M}$, which begs to ask if this is physically grounded. Motivated by current effective one-body models, this work models compact binary coalescence (CBC) as a rotating, compact mass shell that is contracting towards the total mass horizon. Using a variational methodology, the Laplace-Beltrami formulation for the Ricci tensor is applied to a Kerr metric Ansatz, retrieving the energy density $T_{00}$ of the CB mass shell via the Einstein field equations. At the time of coalescence $t_C$, the corresponding surface energy ultimately depends on the reduced mass $\mu$ of the CB, the symmetric mass ratio $\alpha$, and the CB's normalized orbital spin velocity. In other words, this surface energy is the anticipated energy radiated as GWs, which is not one-tenth of the chirp mass systematically. Under simple assumptions, the anticipated energy for GW150914 -- a representative example -- is $2.08 M_\odot c^2$ using documented center values. Under a more rigorous analysis in comparison, the anticipated energy for GW150914 is $3.27M_\odot c^2$. This is compared with the GWTC recorded value of $3.1^{+0.4}_{-0.4}M_\odot c^2$ for GW150914, with the latter analysis providing a closer approximation to the actual value. This study also includes the derivation of gravitational waveforms from the CB mass shell model, which depend on dynamic frequencies and decreasing CB separations.
\end{abstract}

%
%
\submitto{\CQG}
%
\maketitle
%
%

\section{Introduction} \label{sec1}

On September 14, 2015, laser interferometers at LIGO detected a gravitational wave (GW) signal from a binary black hole merger \cite{LIGOScientific:2016aoc}. Bayesian analysis revealed the sources' parameters: the initial total mass was $M\simeq(29+35)M_\odot$ -- calculating a chirp mass $\mathcal{M}=\alpha^{3/5}M$ to be $\sim28M_\odot$ --, and the remnant mass was $M_f\simeq62M_\odot$. Here, $\alpha=m_1m_2/M^2$ is the symmetric mass ratio, and $M_\odot\simeq 2\times10^{30}$ kg defines the solar mass. With the remainder of the initial total mass from the remnant mass converted into the radiated GW energy, this implies $E_{\mathrm{GW}}\simeq2M_\odot$ with $c=1$. At first glance, this shows that $E_{\mathrm{GW}}$ roughly approximates to a tenth of the chirp mass, as the final pulse of GW formation takes place at the chirp phase of coalescence. This pattern continues into further GW detections, regardless of the compact binary (CB) type (see e.g. the GWOSC database \cite{GWOSC} and the comparison between $M-M_f$ and $\sim\mathcal{M}/10$ listed in Refs. \cite{LIGOScientific:2018mvr, LIGOScientific:2021usb, KAGRA:2021vkt}). Although $E_{\mathrm{GW}}\approx\mathcal{M}/10$ is ad hoc, there is an existing correlation to explore, which poses a logical question if it is a physically grounded rule-of-thumb.

GW sources are well understood to be quadrupolar. From the linearized Einstein field equations (EFEs) under a weak field approximation and a compact two-point-mass quadrupole source, one can yield the spatial traceless-transverse (TT) gauged GW profile to assume a simple waveform for circular orbits: 
\begin{equation}\label{gwsourced}
 h^\mathrm{TT}_{ij}=-\frac{4G\mu L^2\Omega^2}{D}\exp(i\,2\Omega t)\varepsilon^\mathrm{TT}_{ij},
\end{equation} 
where $D$ is the luminous distance between the observer and the source, $\mu=m_1m_2/M$ is the reduced mass, $L$ is the CB separation, $\Omega$ is the CB's orbital frequency,  $h_+\propto\cos(2\Omega t)$ and $h_\times\propto\sin(2\Omega t)$ via the $3\times3$ TT-gauged matrix $\varepsilon^\mathrm{TT}_{ij}$, and the strain amplitude is distinctly defined in source-dependent values.

However, the simple wave profile for GW signals is, namely, a simplification for nearly circular orbits in the earliest phases of CB coalescence (CBC), which are subject to post-Newtonian (PN) corrections \cite{Blanchet:2013haa}. Past the PN regime, CBs undergo the inspiral-merger-ringdown (IMR) process where the waveform intensifies with a dynamic frequency and amplitude enhancement, until reaching the maximum peak at the coalescence time $t_C$. Past coalescence, ignoring tidal deformations, the waveform dampens and stiffens into a zero flat-line.

CBC waveforms that depend on numerical relativity, such as the up-to-date \texttt{SEOBNRv4} \cite{Bohe:2016gbl}, the phenomenological \texttt{IMRPhenom} (e.g. \cite{Ajith:2007qp}), and the \texttt{SEOBNRv4T} tidal deformation extension \cite{Hinderer:2016eia, Steinhoff:2016rfi, Steinhoff:2021dsn} among many more, involve systematic models that effectively describe CBC. E.g., the numerical models \texttt{SEOBNRv4} and \texttt{SEOBNRv4T} are integrated around the analytical effective one-body (EOB) model \cite{Buonanno:1998gg}. This model depicts CBC as the CB's reduced mass $\mu$ spiraling into a total mass innermost stable circular orbital (ISCO) radius, before plunging into the horizon with radius $r_S=2GM$. The measure of the symmetric mass ratio $\alpha=\mu/M$ influences the duration of the nearly-circular inspiral path inversely, i.e. how easily nearly-circular orbits collapse into inspiraling plunge paths  \cite{Buonanno:1998gg}. These are reflected in the \texttt{SEOBNRv4} waveform, where nearly-circular orbital waves dominate in earlier times $t<t_C$, before the chirp rise ($t\sim t_C$) and the ringdown ($t>t_C$) in the IMR process respectively enhances and dampens the waveform. These regimes are pieced together by quasi-normal modes, e.g. the ringdown piece incorporated the linearized perturbations from stellar collapse and explosions, derived by Gundlach, Price and Pullin \cite{Gundlach:1993tp, Buonanno:2005xu}.

This work proposes an alternative model for CBC, which is similar in spirit to the EOB model. Instead of considering the reduced mass spiraling into the total mass ISCO radius and plunging into its horizon, we model the binary masses $m_1$ and $m_2$ as distributed along a ring on the orbital plane. Inclination angles $\iota\in[0,\pi]$ are ``integrated'' over by distributing the equatorial mass ring into a hollow mass shell. In this picture, CBC is modeled as a mass shell, whose diameter $S(t)$ is shrinking with the CB separation $L(t)$. The mass measure of this shell is the reduced mass $\mu$: a choice not only logical for binary systems, but one made to be in alignment with the existing and successful EOB model. One can view this mass shell CBC model as a reinterpretation of EOB, which enables us to assume CBs effectively as a singular compact object. Therefore, we can make claims and assumptions consistent with previous EOB and CB approaches, most importantly $D\gg L$. The mass shell, when viewed up-close, has a contracting time-dependent volume $V(t)\propto(S/2)^3$, and is axially rotating to mimic the CB's normalized orbital rotations. The mass shell model and the gravitational waveforms it generates are constructed and discussed in Section \ref{model}.

Across all stages of CBC up to merger, this mass shell approach also upholds Newton's two postulates for a hollow mass shell: respectively, the gravitational field outside the shell is equivalent to that of a point mass, and the gravitational field in the interior is nullified. This treats the generated GWs and their respective waveform as radiation emitted along the shell surface. Using the Einstein field equations (EFEs) $G_{\mu\nu}=8\pi G\,T_{\mu\nu}$, one can isolate and define the energy density $T_{00}\equiv\epsilon$ by determining the curvature encoded in $G_{00}$. Therefore, the coupling between the energy density and the shell volume at the radius $\rho(t)=S(t)/2$ yields the respective surface energy: $E=T_{00}V$. This energy, by this logic, is the anticipated GW energy to test and compare with previous detections and known values. How the energy density, and the surface energy, are obtained and defined is provided in Section \ref{sect:endens}.  After a discussion on alternative analytical routes for deriving the energy density, as well as on possible intrinsic and environmental effects on the CBC mass-shell model, in Section \ref{sect:disc}, we conclude in Section \ref{sect:concl} with a wide scope on how this work may be extended.

\section{Shrinking Shell Model} \label{model}

In stable classical binaries, the reduced mass $\mu$ simplifies two-mass dynamics into a singular system. This is logically no different for CBs, which themselves can be assumed to be a singular object when viewed from a very large luminosity distance $D$. For CBC, the rate of change in the separation $-\dot{L}$ into a final separation $L(t_C)$ (e.g., the total mass horizon diameter in the EOB model \cite{Buonanno:1998gg}) can be viewed up-close as a contracting shell diameter with the rate of change $-\dot{S}$ and a final diameter measure $S(t_C)$. The overdot resembles a time differential: $^\bullet\equiv d/dt$. When integrated over the timelapse $t'\in[t,\,t_C]$, where $t$ is dynamic and $t_C$ is fixed, we yield:
\begin{equation}\label{seps}
-S(t_C)+S(t)=-L(t_C)+L(t).
\end{equation}
In Eq. (\ref{seps}), the CB separation and shell diameter at the coalescence time $t_C$ are fixed values. In this mass shell model, we adapt EOB convention by supposing CBC ends when the masses touch surfaces: $L(t_C)=r_1+r_2$, under the shell diameter belonging to the total mass horizon: $S(t_C)=4GM$. We define, therefore:
\begin{equation}\label{rads}
S(t)=L(t)-r_1-r_2+4GM.
\end{equation}
As previously stated, this mass shell approximation intentionally reinterprets the EOB model, such that the viewpoint is now a shrinking mass shell with the constant measure $\mu$, until reaching the ``innermost'' shell that is the total mass horizon. The radii of the objects influence the calculation of the contracting radius $\rho(t)=S(t)/2$, ensuring a dependence on CB type. 

\subsection{Waveform Model}

Using binary black holes (BBHs) as a representative example, $S(t)=L(t)$ conveniently via Eq. (\ref{rads}) and the BH radii being proportional to their masses $r_i=2Gm_i$. Finding the respective TT-gauged spatial waveform of this mass shell model is conventionally quadrupolar:
\begin{equation}\label{wave}
h^\mathrm{TT}_{ij}=\frac{2G}{D}\ddot{Q}_{ij}^\mathrm{TT},
\end{equation}
where $Q_{ij}$ is the quadrupole moment. From a far-away perspective, the CB mass-shell can be assumed to be a point mass with constant measure $\mu$ and shrinking radius $\rho=L(t)/2$, that is rotating at the CB's dynamic orbital velocity $\Omega=\Omega(t)$. Therefore, the quadrupole moment is proportional to the point mass' moment of inertia $I=\mu \rho^2$:
\numparts
\begin{eqnarray}\label{qij}
& Q_{ij}=\frac{1}{4}\mu L^2\left(c_ic_j-\frac{1}{3}\delta_{ij}\right),\\
&\mathrm{with}\quad\vec{c}=(\cos(\Omega t),\,\sin(\Omega t),0).\label{vecs}
\end{eqnarray}
\endnumparts
Therefore, via Eq. (\ref{wave}) and using Eqs. (\ref{qij}) and (\ref{vecs}), we yield the mass-shell source GW profile readily in the TT-gauge, also considering time variation in the CB separation $L=L(t)$ and angular velocity $\Omega$:
\numparts
 \begin{eqnarray}\label{hplus}
&h_{+}^{\mathrm{TT}}=-\frac{G\mu L^2}{D}\Big(\sin(2\Omega t)\left(\frac{2\dot{L}}{L}\left(\Omega +t\dot{\Omega}\right)+\dot{\Omega}+\frac{t}{2}\ddot{\Omega}\right)\\\nonumber
&\quad\quad\quad\quad\quad\quad\quad -\cos(2\Omega t)\left(\frac{\dot{L}^2}{2L^2}+\frac{\ddot{L}}{2L}-\left(\Omega+t\dot{\Omega}\right)^2\right)+\frac{\dot{L}^2}{2L^2}+\frac{\ddot{L}}{2L}\Big),\\ \label{hcross}
&h_{\times}^\mathrm{TT}=-\frac{G\mu L^2}{D}\Big(\sin(2\Omega t)\left(\frac{\dot{L}^2}{2L^2}+\frac{\ddot{L}}{2L}-\left(\Omega+t\dot{\Omega}\right)^2\right)\\\nonumber
&\quad\quad\quad\quad\quad\quad\quad +\cos(2\Omega t)\left(\frac{2\dot{L}}{L}\left(\Omega +t\dot{\Omega}\right)+\dot{\Omega} +\frac{t}{2}\ddot{\Omega}\right)\Big).
\end{eqnarray}
\endnumparts
If this were repeated for binary neutron stars (BNS) or a BH-NS binary, or for a more generalized approach, one replaces $L\rightarrow S$. If we assume circular orbits, where all rates of change are zero, we recover the wave profiles first seen in Eq. (\ref{gwsourced}), however scaled by $1/4$ due to our choice of quadrupole geometry. To restore the factor of $4$, one can instead re-use the compact two-point-mass geometry for the quadrupole moment, which bypasses the moment of inertia in favor of $m_1r_1^2+m_2r_2^2=\mu L^2$.

 However, unlike the simple circular wave model where $L$ and $\Omega$ are fixed, the quantities are dynamic across CBC while conserving angular momentum $J=I\Omega$. Imposing the conservation law $\dot{J}=0$ yields time-differential expressions that relate the rates of change in $L$ and $\Omega$ to each other:
\begin{equation}
\frac{\dot{\Omega}}{\Omega}=-\frac{2\dot{L}}{L},\quad\quad\frac{\ddot{\Omega}}{\Omega}-\frac{\dot{\Omega}^2}{\Omega^2}=2\left(\frac{\dot{L}^2}{L^2}-\frac{\ddot{L}}{L} \right)\quad\Rightarrow\quad \frac{\ddot{L}}{L}=\frac{\dot{\Omega}^2}{4\Omega^2}-\frac{\ddot{\Omega}}{2\Omega}.
\end{equation}
We see here that the rate of change for the shrinking separation $-\dot{L}$ demonstrates a rate of change where orbital rotations increase. These relations can be put in Eqs. (\ref{hplus}) and (\ref{hcross}) to simplify the waveform expressions, such that the parentheses terms completely consist in rates of change in $\Omega$:
\numparts
 \begin{eqnarray}\label{hp2}
&h_{+}^{\mathrm{TT}}=-\frac{G\mu L^2}{D}\Big(\sin(2\Omega t)\left(-\frac{\dot{\Omega}}{\Omega}\left(\Omega +t\dot{\Omega}\right)+\dot{\Omega}+\frac{t}{2}\ddot{\Omega}\right)\\\nonumber
&\quad\quad\quad\quad\quad\quad\quad -\cos(2\Omega t)\left(-\frac{\ddot{\Omega}}{4\Omega}+\frac{\dot{\Omega}^2}{4\Omega^2}-\left(\Omega+t\dot{\Omega}\right)^2\right)-\frac{\ddot{\Omega}}{4\Omega}+\frac{\dot{\Omega}^2}{4\Omega^2}\Big),\\\label{hc2}
&h_{\times}^\mathrm{TT}=-\frac{G\mu L^2}{D}\Big(\sin(2\Omega t)\left(-\frac{\ddot{\Omega}}{4\Omega}+\frac{\dot{\Omega}^2}{4\Omega^2}-\left(\Omega+t\dot{\Omega}\right)^2\right)\\\nonumber
&\quad\quad\quad\quad\quad\quad\quad +\cos(2\Omega t)\left(-\frac{\dot{\Omega}}{\Omega}\left(\Omega +t\dot{\Omega}\right)+\dot{\Omega} +\frac{t}{2}\ddot{\Omega}\right)\Big).
\end{eqnarray}
\endnumparts

Given the orbital expressions coupled to the wave profiles, which influences the evolution of the wave envelope given CBC dynamics, one might be tempted to simplify the waveforms further. One temptation is to, e.g., constrain $\ddot{\Omega}\rightarrow0$, which is valid for nearly-circular orbits in the early stages of CBC (i.e., when there is no strong angular acceleration). However, for later stages at times $t\sim t_C$, we cannot prematurely discard it if angular acceleration is present. 

For the rest of the study, Eqs. (\ref{hp2}) and (\ref{hc2}) play no further role in obtaining the GW energy released at coalescence, for in Section \ref{sect:endens} a somewhat heuristic approach is utilized to solve for this via a variational methodology on the EFEs. Nonetheless, as the energy density (and the energy flux density) of the GWs scales conventionally by $h_+^2+h_\times ^2$, one can make a convention-centered yet alternative approach to obtain a GW energy expression via Eqs. (\ref{hp2}) and (\ref{hc2}).

\section{Defining the Energy Density} \label{sect:endens}

The energy-momentum tensor $T_{\mu\nu}$ stands on the right-hand side of the EFEs: $G_{\mu\nu}={8\pi G}\,T_{\mu\nu}$. Here, $G_{\mu\nu}$ is the Einstein tensor that is constructed by the metric and Ricci curvature tensors as $R_{\mu\nu}-Rg_{\mu\nu}/2$; the additive cosmological contribution $\Lambda g_{\mu\nu}$ is neglected. As our CB model resembles a spinning mass shell with measure $\mu$ and shrinking radius $\rho(t)=S(t)/2$ via Eq. (\ref{rads}), defining $T_{\mu\nu}$ from the EFEs is direct:
\begin{equation} \label{tens1}
T_{\mu\nu}=\frac{1}{8\pi G}G_{\mu\nu},
\end{equation}
with energy density defined as $\epsilon\equiv T_{00}$. Determining this value, in turn, depends on the geometric information encoded in $G_{00}$, which furthermore depends on $R_{00}$, $R=g^{\mu\nu}R_{\mu\nu}$, relevant Christoffel symbols, and the metric tensor $g_{\mu\nu}$ let alone $g_{00}$. 

One computes $G_{\mu\nu}$ generally ($G_{00}$ specifically) by directly following the standard hierarchy in the EFEs (i.e. following the chain $g_{\mu\nu}\rightarrow g^{\mu\nu}\rightarrow\Gamma^\alpha_{\mu\nu}\rightarrow R_{\mu\nu}\rightarrow R\rightarrow G_{\mu\nu}$ meticulously). Known solutions to the EFEs, e.g. Schwarzschild \cite{Schwarzschild:1916uq, Schwarzschild:1916ae}, Tolman-Oppenheimer-Volkoff \cite{Tolman:1939jz, Oppenheimer:1939ne}, and Kerr \cite{Kerr:1963ud, Boyer:1967, Chandrasekhar:1985kt}, were obtained via this meticulous chain by first assigning a well-defined energy-momentum source for $T_{\mu\nu}$: either vacuum $T_{\mu\nu}=0$ \cite{Schwarzschild:1916uq, Schwarzschild:1916ae, Kerr:1963ud, Boyer:1967} or perfect fluid $T_{\mu\nu}=(\epsilon+p)u_\mu u_\nu+pg_{\mu\nu}$ \cite{Tolman:1939jz, Oppenheimer:1939ne}. However, for a rotating hollow shell, constructing a custom metric from the grass-roots level and carrying out this full procedure would be a cumbersome task; this is more so given that the objective is to solve for $T_{00}$ uniquely for the CB mass-shell model. A more effective alternative is to employ a method analogous to the variational method: beginning with an Ansatz metric $g_{\mu\nu}$ that is a known solution to the EFEs, we apply differential operations to extract components of $G_{\mu\nu}$ and thus obtain effective expressions for the $T_{\mu\nu}$ components.

\subsection{A Variational Equivalent}

In perturbative quantum mechanics, the variational method estimates the energy eigenvalue of an otherwise intractable Hamiltonian by assuming an Ansatz wavefunction, computing the expectation value of the Hamiltonian, and minimizing it to approximate the energy at the ground state. In our context of CBC modeled as a rotating and contracting mass shell, with an unknown but desired energy density, we adopt a similar philosophy. We begin with a well-defined metric with a non-zero determinant as our Ansatz, and apply differential operations that are inherent in the Christoffel symbols and the Ricci tensor. Using the EFEs, we construct the associated energy-momentum tensor $T_{\mu\nu}$ component-wise, such as $T_{00}$.

Since the Ricci tensor schematically obeys $R\sim\partial\Gamma+\Gamma\Gamma$ and the Christoffel symbols satisfy $\Gamma\sim g^{-1}\partial g$, we claim that the Ricci tensor behaves qualitatively like a Laplacian operator acting on the metric tensor: $R\sim \nabla(g^{-1}\partial g)\sim\nabla^2 g$, where $\nabla\sim\partial+\Gamma$ is the covariant derivative. This interpretation is consistent with Lemma 3.32 from Chow and Knopf \cite{Chow:2004}, which expresses the Ricci tensor as a covariant Laplacian:
\begin{equation}\label{ricci}
R_{\mu\nu}\simeq-\frac{1}{2}\nabla_\alpha\nabla^\alpha g_{\mu\nu}\,\left(+~\mathrm{lower~order~terms}\right).
\end{equation}
The operator $\nabla_\alpha\nabla^\alpha$ is the Laplace-Beltrami operator, and when the metric is well-defined and has a non-zero determinant, it can be written in a coordinate expression (i.e. in the Christoffel symbol-free form) as:
\begin{equation}
\nabla_\alpha\nabla^\alpha:=\Delta^\mathrm{LB}=\frac{1}{\sqrt{-g}}\,\partial_\alpha\left(\sqrt{-g}g^{\alpha\beta}\partial_\beta \right),
\end{equation}
where $\sqrt{-g}=\sqrt{-\mathrm{det}(g_{\mu\nu})}$ and $g^{\mu\nu}$ is the inverse metric. By implementing $R_{\mu\nu}=-\Delta^\mathrm{LB}g_{\mu\nu}/2$ (i.e., neglecting the lower-order terms) into the Einstein tensor and neglecting $\Lambda$, we arrive at a Laplace-Beltrami-inspired approximation of $G_{\mu\nu}$ under a given metric Ansatz:
\begin{equation}\label{eins}
G_{\mu\nu}\simeq -\frac{1}{2}\Delta^\mathrm{LB} g_{\mu\nu} -\frac{1}{2}Rg_{\mu\nu} .
\end{equation}
A sanity check is to see whether the Bianchi identities $\nabla^\mu G_{\mu\nu}=0$, and therewith the energy-momentum conservation $\nabla^\mu T_{\mu\nu}=0$, hold in this Laplace-Beltrami formulation, independent on Ansatz metric and choice of coordinates. This is provided in \ref{ap1}, whereby the lower-order terms are considered and explicitly defined.

One observes from Eq. (\ref{eins}) that $G_{00}\propto g_{00}$, with the Laplace-Beltrami operator and Ricci scalar specifically defined given our choice of Ansatz metric. For a CB viewed from a luminosity distance $D\gg S$, we assume it to be a spinning, compact object. Therefore, our choice of Ansatz metric is the Kerr metric \cite{Kerr:1963ud, Boyer:1967, Chandrasekhar:1985kt}, written here in Boyer-Lindquist coordinates for a given rotating BH of mass $m$ as 
\begin{eqnarray}
ds^2=&-\left(1-\frac{2Gm r}{\Sigma} \right)dt^2+\frac{\Sigma}{\Delta}dr^2+\Sigma d\theta^2\\\nonumber
&+\left(r^2+a^2+\frac{2Gm ra^2}{\Sigma}\sin^2\theta \right)\sin^2\theta d\varphi^2-\frac{4Gm ra\sin^2\theta}{\Sigma} d\varphi dt,
\end{eqnarray}
where $a=J/m$ is a length-scale spin parameter, and
\numparts
\begin{eqnarray}
&\Sigma=r^2+a^2\cos^2\theta,\\
&\Delta=r^2-2Gm r+a^2.
\end{eqnarray}
\endnumparts
To modify the Kerr metric so that it reflects the CB mass-shell model, we simply assert that the metric's mass measure is the reduced mass:  $m\rightarrow\mu$.

The metric determinant and Ricci scalar under the Kerr metric are respectively $\mathrm{det}(g_{\mu\nu})=-\Sigma^2\sin^2\theta$ and $R=0$. As expected, the Kerr metric is a vacuum solution for rotating BHs, from which $R_{\mu\nu}=0$. However, under the Laplace-Beltrami interpretation of the Ricci tensor via Eq. (\ref{ricci}), the nontrivial coordinate dependence of the metric allows for non-zero values in $R_{\mu\nu}$, particularly in the $R_{00}$ component. This perspective is particularly useful for determining our unknown energy density for the CB mass shell. Conveniently, all metric elements depend only on $r$ and $\theta$; the Kerr-metric Laplace-Beltrami operator takes a compact form:
\begin{equation}\label{lbk}
\Delta^\mathrm{LB}_\mathrm{Kerr}=\frac{1}{\Sigma \sin\theta}\left[ \sin\theta \partial_r\left(\Delta  \partial_r\right)+\partial_\theta\left(\sin\theta \partial_\theta\right) \right].
\end{equation}
Using Eq. (\ref{ricci}), one can straightforwardly compute $R_{00}$ to be non-zero:
\begin{eqnarray}\label{r00}
R_{00}=\frac{G^2\mu^2}{2\Sigma^4}\Big(&a^4\left(\frac{3}{2}+\frac{1}{2}\cos(4\theta)\right)\\\nonumber
&+4a^2\left(\cos(2\theta)\left(\frac{a^2}{2}-3r^2\right)-3r^2\right)+4r^4 \Big).
\end{eqnarray}

In approaching the Ricci scalar $R$, we can either: (i) set $R=0$ in accordance to the Kerr metric Ansatz, or (ii) heuristically construct $R$ using a Laplace-Beltrami trace over the metric: $R\propto g^{\mu\nu}\Delta^\mathrm{LB} g_{\mu\nu}$. To propose another alternative, motivated by $R=0$ via convention of the Kerr metric: we can define an effective Ricci scalar by leveraging the curvature content of the nonvanishing Kretschmann scalar: $K=R_{\alpha\beta\mu\nu}R^{\alpha\beta\mu\nu}$ \cite{dInverno:1992gxs}.

If we adopt $R=0$, following convention of the Kerr metric as our used metric Ansatz, then the energy density contribution $T_{00}\propto G_{00}$ depends entirely on the Ricci tensor component $R_{00}$. In this work, we take advantage of this useful simplification -- a choice ultimately motivatied out of heuristics. On the other hand, this is a physically naive choice, given that Eq. (\ref{ricci}) can produce non-zero Ricci tensor entries for the Kerr metric Ansatz beyond $R_{00}$. It is furthermore instructive to consider how the full Einstein tensor might be expressed under a non-zero Ricci scalar. In the third case, to be specific, we propose a naive substitution of the Ricci scalar by the negative square-root of the Kretschmann scalar: $R\sim-\sqrt{K}$. This is motivated by dimensional consistency and by energetic considerations (see footnote\footnote{The square-root of $K$ restores dimensional consistency with $R$. The minus sign is introduced heuristically: (i) to perserve the negative $R^2$ contribution in the Kretschmann scalar decomposition: $K=C_{\alpha\beta\mu\nu}C^{\alpha\beta\mu\nu}+2R_{\alpha\beta}R^{\alpha\beta}-R^2/3$, and (ii) to offset the negativity of $g_{00}$ in $G_{00}$, thereby yielding a physically reasonable, positive energy density.}). 

The Ricci scalar $R= g^{\mu\nu}R_{\mu\nu}$, evaluated using the Laplace-Beltrami formulation under the Kerr metric, yields the following explicit result:
\begin{eqnarray}\label{rscale}
&R=2\Big(\frac{64G\mu r^3}{(a^2(1+\cos(2\theta))+2r^2)^3}-\frac{8(a^2+3G\mu r)}{(a^2(1+\cos(2\theta))+2r^2)^2}\\\nonumber
&-\frac{2r^2-4G\mu r+G^2\mu^2}{G\mu r(a^2(1+\cos(2\theta))+2r^2-2G\mu r)}-\frac{\csc^2\theta}{r^2+a^2}\\\nonumber
&+\frac{r\left(a^2+r^2\right)\left(a^2+3r^2-6G^2\mu^2 \right)+2G\mu(a^4-2r^4+3a^2r^2)}{r\left(a^2+r^2\right)\Delta(a^2(1+\cos(2\theta))+2r^2)} \\\nonumber
&+ \frac{(2 r^2 (a^2 + r^2)^2 - (a^2-3r^2)  ( G^2\mu^2 (r^2-a^2)-2 G \mu r  (r^2+a^2)))}{(G\mu r (a^2 + r^2)  (a^4 + 2 r^4 + a^2 r (3 r + 2 G \mu) +    a^2\Delta \cos(2\theta)))}\Big).
\end{eqnarray}
In Eq. (\ref{rscale}), one can observe that the deceptively simple term $-\csc^2\theta/(r^2+a^2)$, originating from the $g^{33}R_{33}$ contribution of the trace, diverges at the poles ($\theta=0,\pi$). This presents a fundamental obstacle, as the CB mass shell was constructed by symmetrically distributing the equatorial mass ring over all inclination angles. In this case, $\iota\in[0,\pi]$ effectively becomes the polar angle domain of the CB mass shell. Therefore, any physical quantity intended for full-sphere integration must remain finite across all $\theta$; this includes the Ricci scalar, which otherwise diverges under integration. This can be side-stepped by imposing $R=0$, reverting to the classical Kerr vacuum, or by selectively omitting the divergent term, ``sweeping it under the rug''. Both options are mathematically expedient, but physically nonsensical.

For the Kerr metric, the Kretschmann scalar is non-zero \cite{Henry:1999rm, Visser:2007fj} (here, utilizing the reduced mass for the mass measure):
\begin{equation}
K=\frac{48G^2\mu^2}{\Sigma^6}\left(r^6+15a^4r^2\cos^4\theta-15a^2r^4\cos^2\theta-a^6\cos^6\theta \right);
\end{equation}
proposing a heuristic modification where $R_\mathrm{eff}=-\sqrt{K}$ allows us to explore the curvature-induced contributions to $G_{00}=R_{00}-R_\mathrm{eff}g_{00}/2$, and how they play a role in calculating the energy density $T_{00}$. This substitution leads to a polynomial in orders of $G$ (a post-Minkowskian (PM) expansion \cite{Damour:2016gwp}):
\begin{eqnarray}
G_{00}\simeq&\frac{G^2\mu^2}{2\Sigma^4}\left(4r^4+4a^2(\dots)+a^4(\dots) \right)\\\nonumber
&-\sqrt{3(r^6+(\dots)-a^6\cos^6\theta)}\left(\frac{2G\mu}{\Sigma^3}-\frac{4G^2\mu^2r}{\Sigma^4} \right),
\end{eqnarray}
where higher order angular terms in $a^2$ and $a^4$, and the square root terms, have been abbreviated for brevity. It is useful to note that the above expression holds for all values of the radial coordinate $r\in(0,\infty)$, provided that the Kretschmann scalar $K$, $R_{00}$ via Eq. (\ref{r00}), and $g_{00}$ via the Kerr metric are general for all $r$. However, the more physically-relevant range of the radial coordinate lies in $r\in[\rho,\infty)$, where $\rho$ is the surface radius of the CB mass shell.

The highest PM order is 2, which can be associated with a generalized gravitational field squared $\vec{\Gamma}^2:=G^2\mu^2/\Sigma^2$. Considering only the 2PM terms, we yield an expression for $G_{00}$ that contains zeroth-order and higher-order spin terms: 
\begin{eqnarray}\label{2pm}
G_{00}^\mathrm{2PM}\simeq\vec{\Gamma}^2\Big[&\frac{1}{2\Sigma^2}\left(4r^4+4a^2(\dots)+a^4(\dots) \right)\\\nonumber
&+\left(\frac{4r}{\Sigma^2} \right)\sqrt{3(r^6+(\dots)-a^6\cos^6\theta)}\Big].
\end{eqnarray}

\subsection{Expressing the Energy}

Here, we express the energy density $T_{00}$ using the simplified relation $G_{00}\approx R_{00}$ via Eq. (\ref{r00}) and $R=0$. Integrating the polar angle over the full range $\theta\in[0,\pi]$, we obtain
\begin{equation}\label{roo}
R_{00}\Rightarrow \frac{\pi G^2\mu^2r(2r^2-3a^2)}{(r^2+a^2)^{7/2}}.
\end{equation}
A kinematic regime of particular relevance for CBC is the low-spin limit. While CBC leads to pre-merger orbital velocities in the relativistic regime, the final spin parameter is typically sub-extermal with $a<1$ \cite{LIGOScientific:2018mvr, LIGOScientific:2021usb, KAGRA:2021vkt}. That is, across all stages of CBC, the spin parameter remains well below unity. This physical constraint motivates a Taylor expansion of Eq. (\ref{roo}) in small $a$. Expanding up to quadratic order reveals a zeroth-order Newtonian term and a second-order spin correction:
\begin{equation}\label{roos}
\Rightarrow R_{00}=\frac{2\pi G^2\mu^2}{r^4}-\frac{10\pi G^2\mu^2}{r^4}\frac{a^2}{r^2}+\mathcal{O}(a^3).
\end{equation} 
Therefore, the energy density of the CB mass shell model is
\begin{equation} \label{endens1}
T_{00}\simeq \frac{G\mu^2}{4r^4}\left(1-5\frac{a^2}{r^2} \right);
\end{equation}
on the CB mass shell at $r=\rho$, the surface energy $E=T_{00}V$ calculates to be
\begin{equation} \label{energy}
E\simeq \frac{\pi}{3}\frac{G\mu^2}{\rho}\left(1-5\frac{a^2}{\rho^2} \right).
\end{equation}

The spin parameter $a=J/\mu$ encodes the angular momentum, which depends on the moment of inertia $J=I\Omega$. As like before for Eq. (\ref{qij}) for the quadrupole moment, when viewed from the luminosity distance $D\gg\rho$, the compact hollow mass shell can be assumed to be a point mass with $I=\mu\rho^2$. This calculates the spin parameter $a=\rho^2\Omega$, which is the coupling $\rho v$ where $v$ is the normalized rotational velocity of the CB. Recovering $c$, the speed ratio $\beta=v/c$ will be utilized in the spinning contribution. Therefore, at the time of coalescence $t_C$, where respectively $\rho(t_C)=2GM$ and $\beta_C=v_C/c$, we yield the energy at the peak of the chirp rise in the CBC waveform, with $c=1$:
\begin{equation}\label{erad}
E(t_C)\simeq \frac{\pi}{6}\frac{\mu^2}{M}\left(1-5\beta_C^2\right)=\frac{\pi}{6}\alpha\mu(1-5\beta_C^2).
\end{equation}
In Eq. (\ref{erad}), the symmetric mass ratio $\alpha=\mu/M$ is recovered, and the energy scaling therefore relies on the interplay of the CB masses, their \textit{reduced} mass, and their normalized spin velocity at merger. This is, by appearance alone, \textit{not} the ad hoc scaling of $E\approx\mathcal{M}/10$. 

The spin factor $(1-5\beta^2)$ in Eq. (\ref{energy}) shows a decrease in energy as $\beta$ increases. Past a certain value of $\beta$, the spin factor becomes negative, which puts the energy into a forbidden zone that we must avoid. Since the rotational velocity increases as CBC progresses, we can constrain a maximum $\beta_C$. The energy becomes zero at the value of $\beta^2=1/5$; implying that this is a ceiling value for the normalized rotational velocity at coalescence, the coalescence velocity ratio must not exceed $\beta_C\leq0.447$, or equivalently, recovering $c$:
\begin{equation}
v_C\leq0.447c.
\end{equation}  
This is compared to the rotational velocity at the total mass ISCO radius, which can serve as an extra constraint:
\begin{equation}
v_\mathrm{ISCO}=\sqrt{\frac{GM}{r_\mathrm{ISCO}}}\simeq 0.408 c.
\end{equation}

A representive example is GW150914, initially introduced in Section \ref{sec1} to have the total mass of $M\simeq(29+35)M_\odot$. This calculates the reduced mass to be $\mu=15.86M_\odot$ and the symmetric mass ratio to be $\alpha=0.248$. We have a ``ballpark'' number for the radiated GW energy to be $E_\mathrm{GW}\simeq 2M_\odot c^2$, with reported values to be $E_\mathrm{GW}=3.0^{+0.5}_{-0.5}M_\odot c^2$ in the detection paper \cite{LIGOScientific:2016aoc} and $E_\mathrm{GW}=3.1^{+0.4}_{-0.4}M_\odot c^2$  in the GWTC paper \cite{LIGOScientific:2018mvr}. Given the peak frequency of the GW being $\sim250$ Hz \cite{LIGOScientific:2016aoc}, we can calculate the anticipated rotational speed of the CB at merger using the total mass horizon radius and half of the GW's frequency: $v_C=0.0791 c$. Using Eq. (\ref{erad}), the anticipated energy radiated from the mass shell surface at time $t_C$ is
\begin{equation} \label{reps}
E(t_C)\simeq 1.994M_\odot c^2.
\end{equation}
Using the up-to-date center values for the CB masses in Refs. \cite{GWOSC, LIGOScientific:2021usb} and the same anticipated rotational speed, the anticipated energy calculates to be $E\simeq2.075M_\odot c^2$. Even for a representative case, these anticipated energy values are close to the ``ballpark'' and reported values, with 1:1 ratio values being $0.996$ ``ballpark'' and roughly $2/3$ based on reported center values. More importantly, they are at the same order of magnitude.

Another example to test is GW170817, which is a GW sourced by BNSs \cite{LIGOScientific:2017vwq}. Unlike BBHs, BNSs exchange tidal deformations that complicate the CBC process by imposing dynamic tides \cite{Hinderer:2016eia} that shift the IMR orbital phase up to the 6PN expansion order \cite{Flanagan:2007ix, Favata:2013rwa, Chatziioannou:2020pqz}. While this is reflected in the waveforms given as Eqs. (\ref{hp2}) and (\ref{hc2}), these effects should be independent of the energy reading via Eq. (\ref{erad}), apart from the normalized rotational velocity at coalescence. Via Refs. \cite{GWOSC, LIGOScientific:2018mvr}, the binary mass measures are recorded to be $m_1=1.46^{+0.12}_{-0.10}M_\odot$ and $m_2=1.27^{+0.09}_{-0.09}M_\odot$ using low spin priors. The center values calculate a reduced mass of $\mu=0.679 M_\odot$ with a symmetric mass ratio $\alpha=0.249$. With the GW peak frequency being $\sim300$ Hz \cite{LIGOScientific:2017vwq}, the anticipated normalized rotational speed at coalescence is calculated to be $v_C=0.00405c$. The radiated GW energy is constrained to be $E\geq 0.04 M_\odot c^2$ \cite{GWOSC, LIGOScientific:2018mvr}; the mass shell model anticipates the energy to be $E(t_C)\simeq0.0885 M_\odot c^2$ via Eq. (\ref{erad}). The chirp mass ad hoc scaling gives another expectation of the GW170817 energy to be $E\sim 0.1186 M_\odot c^2$ via the spin-prior-mutual chirp mass $\mathcal{M}=1.186^{+0.001}_{-0.001}M_\odot$ \cite{GWOSC, LIGOScientific:2018mvr}.

\section{Discussion}\label{sect:disc}

The discrepancy between Eq. (\ref{reps}) and the reported values for GW150914's radiated energy suggests a need to improve the energy density expression. Assuming the variational methodology and the Kerr metric Ansatz collectively remain a valid strategy, the precision in our framework rests on our choice for $G_{00}$. Initially, we adopted $R_{00}$ via Eq. (\ref{r00}) -- Eq. (\ref{roo}) when integrating over the polar angle $\theta$ -- with $R=0$ as a caveat for using the Kerr metric. Since $R\propto g^{\mu\nu}\Delta^\mathrm{LB} g_{\mu\nu}$ via Eq. (\ref{rscale}) is ultimately divergent under a polar-angular integral, we resort to determining the energy density $T_{00}$ and its respective energy $E$ using Eq. (\ref{2pm}) as the corresponding Einstein tensor element. 

The angular integration of Eq. (\ref{2pm}) poses significant difficulty, especially due to the second term involving a square root of angular dependent quantities. While substitution techniques such as the Weierstraß transformation may be applicible for analytical control, a more tractable approach is to perform a small-$a$ Taylor expansion of Eq. (\ref{2pm}) up to quadratic order, and then integrate over $\theta$. This not only simplfies the integration but also serves as a consistency check by verifying that we recover Eq. (\ref{roos}) from this ``reverse-step'' pathway. Carrying out this procedure, we obtain:
\begin{eqnarray}\label{goo}
\Rightarrow G_{00}^\mathrm{2PM}=&\frac{2\pi G^2\mu^2}{r^4}-\frac{10\pi G^2\mu^2}{r^4}\frac{a^2}{r^2}-\frac{2G\mu\sqrt{3}}{r^3}+\frac{21\pi G\mu\sqrt{3}}{r^3}\frac{a^2}{r^2}\\\nonumber
&+\frac{4\pi G^2\mu^2\sqrt{3}}{r^4}-\frac{23\pi G^2\mu^2\sqrt{3}}{r^4}\frac{a^2}{r^2}.
\end{eqnarray}
The first two terms of Eq. (\ref{goo}) exactly reproduce Eq. (\ref{roos}). We also yield from this 2PM Einstein tensor element yet another PM expansion of maximum order 2. Discarding the 1PM terms, Eq. (\ref{goo}) simplifies to a modified zeroth-order Newtonian term and second-order spin correction compared to Eq. (\ref{roos}):
\begin{equation}\label{goos}
G_{00}^\mathrm{2PM}\simeq\frac{8.928\pi G^2\mu^2}{r^4}-\frac{49.837\pi G^2\mu^2}{r^4}\frac{a^2}{r^2}.
\end{equation}
Compared to Eq. (\ref{roos}), both the Newtonian and spin terms are quantitatively enhanced. These enhancements arise directly from the inclusion of Kretschmann-derived terms in the effective Einstein tensor, which would lead to an overestimation of the energy density and CB surface energy. This is demonstrated by the resulting energy density:
\begin{equation} \label{endens2}
T_{00}\simeq 1.116\frac{G\mu^2}{r^4}\left(1-5.582\frac{a^2}{r^2} \right),
\end{equation}
and the respective mass-shell surface energy:
\begin{eqnarray} \label{energy2}
&E\simeq 4.675\frac{G\mu^2}{\rho}\left(1-5.582\beta^2 \right)\\\nonumber
& \Rightarrow\quad E(t_C)=2.337\frac{\mu^2}{M}\left(1-5.582\beta_C^2 \right).
\end{eqnarray}

In Eq. (\ref{energy2}), it is evident that the anticipated values are systematically overestimated compared to the initial formulation, roughly by a factor of $4.5$ from the pre factor alone. Using GW150914 and its parameters as the representative example again, the anticipated energy at time $t_C$, according to the revised mass-shell model, gives the ``ballpark'' number of $E\simeq 8.865 M_\odot c^2$ and $E\simeq 9.227 M_\odot c^2$ via the reported center values. These values severely overshoot the known values, and as expected, this discrepancy suggests that the Kretschmann scalar contributions are overscaled relative to the original $R_{00}$ minimum. 

To remedy this, we propose introducing a reciprocal scaling factor $\lambda$, such that the effective Ricci scalar becomes $R_\mathrm{eff}=-\lambda\sqrt{K}$. To avoid ad hoc parameter fitting, we impose the constraint $\lambda^{-1}\in\mathbb{N}$ and consider values with geometric significance. E.g., choosing $\lambda=1/6$ yields the expected ``ballpark'' GW150914 energy of $E\simeq 3.14 M_\odot c^2$, as well as the energy $E\simeq3.267 M_\odot c^2$ using reported center values. Using the same rescaling to GW170817, the anticipated energy value yields $E\simeq0.140 M_\cdot c^2$, which is still comparable to the ad hoc chirp mass expectation $E\sim 0.1186 M_\odot c^2$. At first glance, this may appear to be a random guess. However, we observe that $\lambda=1/6$ decomposes naturally: $\lambda=1/2\cdot1/3$. The factor $1/2$ matches the same scaling as the Laplace-Beltrami form of the Ricci tensor (c.f Eq. [\ref{ricci}]), which would have persisted in a full treatment of $R$ from variational principles. The factor $1/3$ is therefore the Kretschmann scalar contribution (see footnote\footnote{ The $1/3$ factor emerges from the identity $K=C_{\alpha\beta\mu\nu}C^{\alpha\beta\mu\nu}+2R_{\alpha\beta}R^{\alpha\beta}-\frac{1}{3}R^2$, where the $1/3$ factor appears in the Ricci-scalar-squared term. This identity clarifies that a pure Kretschmann scalar cannot be substituted wholesale for $R$ without acknowledging the scalings of relevant Ricci components.}). This underscores that the correction is not arbitrary, as one might initially assume, but rather a natural consequence of the underlying curvature decomposition.

\subsection{On Eccentric or Non-Circular Orbits}\label{eccent}

It must be noted that the CB mass-shell waveforms, presented as Eqs. (\ref{hp2}) and (\ref{hc2}), inherently assume intensifying, (nearly-)circular orbits up to plunge. This is reflected in the sinusoidal waveform profile and in the time derivatives (up to second order) of the orbital velocity $\Omega$ that shape the wave amplitude envelope. When deviations from circularity, e.g. quasi- and non-circular orbits, are considered, the vectors that enter the quadrupole moment as Eq. (\ref{vecs}) must be redefined to capture these off-circular orbital geometries, e.g. elliptical components in $\vec{c}$:
\begin{equation}
\vec{c}=\frac{2}{L}(r_+\cos(\Omega t)+u,\,r_{-}\sin(\Omega t)+w,0),
\end{equation}
where $r_+$ is the semi-major axis, $r_{-}$ is the semi-minor axis, and the position coordinates $(u,w)$ mark the elliptic center.  We recall that $L/2$ acts as the mass-shell surface radius $\rho$ in the context of Eqs. (\ref{hp2}) and (\ref{hc2}).  Another alternative is the expansion of the orbital frequency in the similar spirit as post-Newtonian expansion \cite{Blanchet:2013haa, Favata:2013rwa}:
\begin{equation}
\Omega(t)=\Omega_\mathrm{circ}(t)\times\left(1 + \delta\right),
\end{equation}
where $\Omega_\mathrm{circ}$ is the circular orbit frequency and $\delta$ is an off-circular deviation that, qualitatively, includes eccentricity $e$ and the eccentric anomaly $E$ -- such as $\delta\sim e\cos E$. Either choice of modification (or both, per one's choice), as a result, affects directly the waveform morphology and, consequently, the temporal evolution of the wave envelope.

A characteristic parameter of quasi-circular orbits is the eccentricity $e$. Its influence on the IMR orbital phase shift appears as the 2PN expansion order \cite{Favata:2013rwa} (1.5PN in Ref. \cite{McMillin:2025hof}). Analogously to the discussion on the 6PN tidal correction in the IMR orbital phase shift, when utilizing GW170817 as an example, these eccentricity-induced terms are not expected to modify the total energy radiated, both according to Eq. (\ref{erad}) and the reduced Kretschmann scalar re-calculation -- that is, except indirectly through the normalized rotational velocity at coalescence. 

A representative example of an eccentricity-rich GW event is GW190521 \cite{LIGOScientific:2020iuh}: a BBH merger with a total mass center value of $153.1 M_\odot$ \cite{GWOSC, LIGOScientific:2021usb}. It has been interpreted through consistency checks as a highly eccentric merger event with the eccentricity $e\approx0.7$ \cite{Gayathri:2020coq}. With the binary masses being $m_1=98.4^{+33.6}_{-21.7}M_\odot$ and $m_2=57.2^{+27.1}_{-30.1}M_\odot$ \cite{LIGOScientific:2021usb}, the center values calculate a reduced mass of $\mu=36.2 M_\odot$ and a symmetric mass ratio $\alpha=0.232$. Given the peak GW frequency is approximately $\sim100$ Hz \cite{GWOSC}, the mass-shell model predicts the total emitted energy of $E(t_C)\simeq4.27 M_\odot c^2$ via Eq. (\ref{erad}) and $E\simeq 6.73 M_\odot c^2$ through the reduced Kretschmann scalar alternative. The cataloged GW energy, inferred from the difference between the total mass and the remnant mass, is $E\simeq 5.7 M_\odot c^2$ through the center values \cite{GWOSC, LIGOScientific:2021usb}. The small discrepancy between this observationally inferred value and the Kretschmann-based estimate (i.e., the energy shift of order $\sim1M_\odot c^2$)  may, after all, suggest eccentricity-induced effects in the GW coalescence energy; this warrants further investigation in future PN-influenced extensions to the model (e.g. eccentricity and tidal deformation).

\subsection{On the CBC Ringdown Phase}

Because the mass-shell model is in spirit a reinterpretation of the EOB framework for CBC, the missing piece for the mass-shell model waveforms via Eqs. (\ref{hp2}) and (\ref{hc2}) is the ringdown phase. In the EOB framework, the ringdown phase is pieced together by quasi-normal modes, e.g. a Gaussian-like damping due to stellar collapse and explosions \cite{Gundlach:1993tp, Buonanno:2005xu}. For the mass-shell model waveforms, this is no different.

In the general sense, using Eqs. (\ref{hp2}) and (\ref{hc2}) as the basis for the inspiral and merger phases, the complete waveform with a pieced-in ringdown phase is essentially a linear expresson of sine and cosine wave profiles, each are scaled by a time-dependent and polarization-specific ``envelope function'' $\mathfrak{E}(t)$ encoding dynamics:
\begin{equation} \label{pieces}
h_{+/\times}^\mathrm{TT}(t)=-\frac{G\mu L^2}{D}\left[\cos(2\Omega t) \mathfrak{E}_{+/\times}^\mathrm{cos}(t)+\sin(2\Omega t)\mathfrak{E}_{+/\times}^\mathrm{sin}(t)\right].
\end{equation}
E.g., for the simple wave approximation where $\dot{\Omega},~\ddot{\Omega}\rightarrow0$, $\mathfrak{E}(t)$ is uniquely defined (Either $\Omega^2$ or 0) for all $t$, such that the wave is only scaled by its amplitude similar to Eq. (\ref{gwsourced}). However, the dynamic amplitude across IMR lies in a characteristic piece-wise functionality in $\mathfrak{E}(t)$, which for times $t\leq t_C$ is the complete series of expressions provided in Eqs. (\ref{hp2}) and (\ref{hc2}). For times $t>t_C$, a Gaussian-like damping motivated by Ref. \cite{Gundlach:1993tp} is pieced after merger, which characteristically results as the level-out to a zero flatline. It is of interest to analyze the envelope function $\mathfrak{E}(t)$ in a future study, especially for BNSs whose waveforms exhibit tidal contributions up to merger and post-merger waveforms before BH collapse. 

\subsection{On Further Mass-Shell Model Improvements}

\subsubsection{Including Component-body Spin and Charge}\label{spch}

The work here centers around an ideal CB whereby the component bodies are characterized only by their masses $m_1$ and $m_2$. For all binary types (BBH, BNS, and BH-NS), component body spins (i.e. individual axial rotations) contribute uniquely to CBC by influencing spin-prior-dependent initial and final state measureables \cite{LIGOScientific:2018mvr, LIGOScientific:2021usb, KAGRA:2021vkt}, and the possibility of charged BBHs has previously been explored (e.g. \cite{Benavides-Gallego:2022dpn, Grilli:2024fds}). How the other no-hair credentials (charge and spin, next to mass) improve the CB mass-shell model and furthermore contribute to the GW coalescense energy rests more in the Laplace-Beltrami formalism than in the waveforms via Eqs. (\ref{hp2}) and (\ref{hc2}).
\begin{enumerate}
\item Because the variational method on the EFEs via the Laplace-Beltrami formalism takes inspiration from quantum mechanical variational method, the treatment of a CB mass shell with spinning binary components would adopt yet another detail of quantum mechanics: the unique property of intrinsic spin (a.k.a, plainly, \textit{spin}). Because the CB mass-shell model uses a Kerr metric Ansatz, the orbital rotation of the binary system is the effective axial spin of the mass shell. However, should the component masses, themselves, have axisymmetric rotations, this is effectively describing an intrinsic rotation -- an analog to intrinsic spin -- of the CB mass shell.\\

In the spirit of rotational quantum mechanics, the total angular momentum of the CB mass shell $\mathcal{J}$ entails the mass shell's rotational angular momentum $J=I\Omega$ and some spin-like intrinsic angular momentum for each binary mass $\mathcal{S}_i$ in the form of $\mathcal{J}=J+\mathcal{S}_1+\mathcal{S}_2$. This influences the spin parameter (for the CB mass shell) $a=J/\mu$ in the Kerr metric Ansatz such that $a\rightarrow\mathcal{J}/\mu$, and given $\mathcal{J}$ is presented to be linear, the Kerr metric spin parameter is also linear: $a\rightarrow J/\mu+\mathcal{S}_1/\mu+\mathcal{S}_2/\mu$. This will impose further correction terms in e.g. Eq. (\ref{energy}) in the form of
\begin{equation}
E\rightarrow \frac{\pi}{3}\frac{G\mu^2}{\rho}\left[1-\frac{5}{\rho^2}\left( \frac{J}{\mu}+\frac{\mathcal{S}_1}{\mu}+\frac{\mathcal{S}_2}{\mu}\right)^2 \right].
\end{equation}
If the binary masses are not spinning ($\mathcal{S}_i\rightarrow0$), we recover Eq. (\ref{energy}) specifically. On the other hand, one can motivate -- however naively -- $\mathcal{J}=\mathcal{I}\tilde{\Omega}$, such that the total angular momentum is effectively defined in the form of a standard rotational angular momentum with a ``moment of inertia" $\mathcal{I}$ and an ``angular frequency" $\tilde{\Omega}$. This motivation intends to enable these intrinsic spin corrections to influence the CB mass-shell waveforms via Eqs. (\ref{hp2}) and (\ref{hc2}) by asserting $\Omega\rightarrow\tilde{\Omega}$. We want to emphasize that this mapping is a qualitative device to incorporate intrinsic spin corrections in the waveform phase and envelope structure; a proper analysis of the waveform-level impact is deferred to future work. \\

\item Suppose each binary mass has an electric charge, $q_1$ and $q_2$. The charge of the binary, much like an ordinary electrostatic system, is a superposition of charges in the form of an enclosed charge $q_\mathrm{encl}=q_1+q_2$. This enclosed charge is also the charge of the CB mass shell, which is distributed in the same manner as the reduced mass: across the shell. The Ansatz metric -- now describing the CB effectively as a compact, rotating, and charged body -- is instead the Kerr-Newman metric \cite{Newman:1965tw, Newman:1965my} (another solution to the EFEs for a charged, spinning mass), which is written as follows in Boyer-Lindquist coordinates for a BH with mass $m$ and charge $\mathcal{Q}$:
\begin{eqnarray}
ds^2=&-\left(1-\frac{2Gm r}{\Sigma}+\frac{\mathcal{Q}^2G}{4\pi\varepsilon_0\Sigma} \right)dt^2+\frac{\Sigma}{\Delta}dr^2+\Sigma d\theta^2\\\nonumber
&+\frac{\chi}{\Sigma}\sin^2\theta d\varphi^2+\frac{2a}{\Sigma}\left(\frac{\mathcal{Q}^2G}{4\pi\varepsilon_0}-2Gm r\right)\sin^2\theta d\varphi dt,
\end{eqnarray}
where $\varepsilon_0$ is the electric permittivity in vacuum, and all Kerr metric parameters are previously defined except for
\numparts
\begin{eqnarray}
&\Delta=r^2-2Gm r+a^2+\frac{\mathcal{Q}^2G}{4\pi\varepsilon_0},\\
&\chi=(a^2+r^2)^2-a^2\sin^2\theta\Delta.
\end{eqnarray}
\endnumparts
Therefore, the combination $\mathcal{Q}^2G/(4\pi\varepsilon_0)$ is a charge-dependent length scale, which naturally vanishes in the pure Kerr case ($\mathcal{Q}\rightarrow0$). For the Kerr-Newman metric to act as the Ansatz metric for a charged CB mass shell, we assert that $m\rightarrow\mu$ and $\mathcal{Q}\rightarrow q_\mathrm{encl}$.\\

The determinant and the Ricci scalar of the Kerr-Newman metric is exactly those for the Kerr metric: $\mathrm{det}(g_{\mu\nu})=-\Sigma^2\sin^2\theta$ and $R=0$. However, the Laplace-Beltrami formalism can provide non-zero expressions for the $R_{\mu\nu}$ entries in the Kerr-Newman case, just as it did for pure Kerr. Therefore, the Kerr-Newman Laplace-Beltrami operator holds the same form:
\begin{equation}
\Delta^\mathrm{LB}_\mathrm{KN}=\frac{1}{\Sigma \sin\theta}\left[ \sin\theta \partial_r\left(\Delta  \partial_r\right)+\partial_\theta\left(\sin\theta \partial_\theta\right) \right],
\end{equation}
while keeping in mind that $\Delta$ now contains an electrostatic contribution. And via $R_{00}\propto\Delta^\mathrm{LB}g_{00}$, the non-zero $R_{00}$ is computed to be the following, readily implying full-sphere integration over $\theta$ and Taylor expansion for small $a$:
\begin{eqnarray}\label{rooel}
\Rightarrow R_{00}=&\frac{2\pi G^2\mu^2}{r^4}-\frac{10\pi G^2\mu^2}{r^4}\frac{a^2}{r^2}+\frac{Gq_\mathrm{encl}^2}{4\varepsilon_0 r^4}-\frac{3G^2\mu\,q_\mathrm{encl}^2}{2\varepsilon_0r^5}\\\nonumber
&+\frac{17G^2\mu\,q_\mathrm{encl}^2}{4\varepsilon_0r^5}\frac{a^2}{r^2}+\frac{3G^2q_\mathrm{encl}^4}{16\pi\varepsilon_0^2r^6}-\frac{13G^2q_\mathrm{encl}^4}{32\pi \varepsilon_0^2r^6}\frac{a^2}{r^2}.
\end{eqnarray}
Under the Kerr-Newman metric Ansatz, the GW energy (if one follows the simple assumption of $R=0$ in alignment with Kerr-Newman convention) has uniquely-defined electrostatic and gravito-electric terms. As Eq. (\ref{rooel}) is presented as a PM expansion up to order 2, the only 1PM term is classically electrostatic if one further defines $T_{00}\approx R_{00}/(8\pi G)$ and therewith $E=T_{00}V$.\\

However, it is worth noting that  realistic astrophysical BHs are not expected to have a significant electric charge, as the matter that forms them is largely electrically neutral. In addition, any charge left over would be quickly neutralized by the surrounding accretion plasma. Nonetheless, should BHs have an electric charge, it is physically constrained to be smaller than the mass measure: $\mathcal{Q}\ll m$. For a CB of such BHs, the electrostatic contributions would be grossly overpowered by the main gravitational contributions to provide any new information, but for completeness these terms are presented in Eq. (\ref{rooel}).
\end{enumerate} 

\subsubsection{Involving Dark and/or Baryonic Matter}

The present work assumes a pure Kerr Ansatz spacetime, in which the rotating mass corresponds to the reduced mass of a CB. Should a CB be in the vicinity of dark matter (DM) and/or baryonic matter (BM), these environments may produce localized perturbations to the CB mass-shell curvature through their respective gravitational signatures. Such perturbations would directly modify the local spacetime geometry and with it an altered GW waveform relative to Eqs. (\ref{hp2}) and (\ref{hc2}). 

E.g., should the CB mass shell be surrounded by a simple baryonic gas (as a representative example) that imposes a form of ``drag force," the time-differential of the angular momentum would be non-zero due to the dragging by the gaseous medium: $\dot{J}=-\gamma J$, where $\gamma$ is a dimensionless dragging coefficient. Thus, the first-order derivative of $\Omega$ and the seperation length $L$ are expressed in a manner that these rates of change depend on this coefficient:
\begin{equation}\label{damps}
\Rightarrow 2\frac{\dot{L}}{L}+\frac{\dot{\Omega}}{\Omega}=-\gamma.
\end{equation}
However, the second-order derivatives would not depend on $\gamma$, as a time-derivative of Eq. (\ref{damps}) on the right-hand side is zero. This recovers the original expression involving terms with $\ddot{\Omega}$ and $\ddot{L}$, as they define the non-linear, Newtonian-like attraction of CB components. However, all first-order terms in Eqs. (\ref{hplus}) and (\ref{hcross}) will include this dragging coefficient as an added term, if one chooses to express the waveforms in the style of Eqs. (\ref{hp2}) and (\ref{hc2}).

Within the Laplace-Beltrami variational framework, environmental effects for DM and/or BM on the coalescence energy can be approached by the following choices of improvement:
\begin{enumerate}
\item For the case of DM, being it is collisionless but gravitational, the metric Ansatz may deviate from pure Kerr into, e.g., a ``Kerr $+$ DM" linearization scheme: $g_{\mu\nu}=g^\mathrm{Kerr}_{\mu\nu}+h^\mathrm{DM}_{\mu\nu}$, where $h^\mathrm{DM}_{\mu\nu}$ is the DM-induced metric, with $g^{\mu\nu}=g_\mathrm{Kerr}^{\mu\nu}-h_\mathrm{DM}^{\mu\nu}$ to motivate $g^{\alpha\mu}g_{\beta\mu}=\delta^\alpha_{~\beta}$. If the DM perturbation has a non-zero determinant, this modification systematically alters the coordinate-dependent Laplace-Beltrami operator:
\begin{eqnarray}
\Delta^\mathrm{LB}g_{\mu\nu}=&\frac{1}{\sqrt{-g-h}}\\\nonumber
&\times\partial_\alpha\left(\sqrt{-g-h}\left(g_\mathrm{Kerr}^{\alpha\beta}\partial_\beta-h_\mathrm{DM}^{\alpha\beta}\partial_\beta\right) \right)\left( g^\mathrm{Kerr}_{\mu\nu}+h^\mathrm{DM}_{\mu\nu}\right).
\end{eqnarray}
Given that $R_{00}\propto\Delta^\mathrm{LB}g_{00}$ in the Laplace-Beltrami formalism, the provided expansion defines unique DM contributions to the GW energy expression. Should one remove the DM contribution ($h^\mathrm{DM}_{\mu\nu}\rightarrow0$), the formulation naturally reduces to the pure Kerr case. As a representative example, one might write the $g_{00}$ component as a linear combination of the Kerr potential and a DM potential (using the reduced mass as the rotating mass in the Kerr metric):
\begin{equation}\label{trial}
g_{00}=-\left(1-\frac{2G\mu r}{\Sigma} +\Phi_\mathrm{DM}\right),
\end{equation}
where $\Phi_\mathrm{DM}$ derives from a chosen DM density profile, e.g. a spike $\rho_\mathrm{DM}\approx\rho_0(r/r_0)^{-\gamma}$ with $1<\gamma<2.5$.\\

 It is worth noting that, alongside Eq. (\ref{trial}), the DM metric and its determinant are needed to be well-defined in order to extract DM-specific contributions to the GW coalescence energy. This is rather complicated if one does not have a well-defined metric without constructing one ad hoc. Thus, one must find a solution for DM by conventionally solving the EFEs. Alternatively, a modified GR framework that readily incorporates DM and dark energy directly in the EFEs \cite{Nash:2023zza} could be adopted while retaining the Laplace-Beltrami formalism to solve for the DM-dressed GW coalescence energy. Select literature that further discuss DM effects on e.g. BBHs are offered as Refs. \cite{Bertone:2024rxe, Chakravarti:2025xaj}, and Ref. \cite{Miller:2025yyx} provides an extensive review of how GWs can act as probes on particle DM.\\

\item For the case of BM, the influence of surrounding gas, plasma, or accreting material is inherently dynamical rather than purely geometric. Instead of perturbing the Kerr metric Ansatz, these dynamic responses by the BM environment can be represented as additional terms in the energy-momentum tensor: $T_{\mu\nu}=T^\mathrm{CB}_{\mu\nu}+T^\mathrm{BM}_{\mu\nu}$. If one follows the pure Kerr Ansatz in the Laplace-Beltrami formalism, the geometric signature in e.g. $G_{00}$ via Eq. (\ref{goo}) (while utilizing the reduced Kretschmann scalar) is instead proportional to $T_{00}^\mathrm{CB}+T_{00}^\mathrm{BM}$. The baryonic energy density thereby acts as an effective damping term on the total GW coalescence energy, consistent with the notion that dense matter environments can absorb or scatter a fraction of the emitted radiation (see footnote\footnote{A further implication of this damping is the ``domino effect" of GWs probing the BM and causing a form of turbulence. This notion of GW-causing BM turbulence was originally suggested out of frutiful discussion with Mr. Percy Martinez, at the Universit\"at Potsdam.}). Such effects are expected to be negligible for stellar-mass binaries in vacuum-like conditions, but could become relevant for systems embedded in high-density environments, such as supermassive BH mergers surrounded by circumbinary disks. 
\end{enumerate}

\subsubsection{Applying Formation Channels}

The formation of CBs is generally attributed to two pathways: isolated binary evolution, and dynamical assembly in dense stellar environments. These channels imprint distinct signatures on pre-merger parameters, such as eccentricity $e$, mass ratio $q:=m_2/m_1$ and the degree of spin alignment of the binary masses. While the CB mass-shell model does not directly infer the formation channel, it readily incorporates the reduced mass, orbital angular momentum, and normalized rotational velocity as fundamental input parameters (see Eq. (\ref{energy}) for the energy in the inspiral and pre-merger phases and Eq. (\ref{erad}) for the coalescence energy). These quantities retain, at least implicitly, the CB's dynamical history and its morphology over time.

E.g., dynamically formed binaries may result to a non-zero eccentricity value or exhibit significant spin misalignment. Within the mass-shell model, these behaviors appear respectively through modifications to the quadrupole structure (see Section \ref{eccent}) or through the angular momentum corrections leading to an adjusted Kerr spin parameter (see Section \ref{spch} (i)). On the other hand, isolated binaries --  expected to circularize efficiently and to possess predominantly aligned component spins -- correspond to more symmetric configurations, which are also presented in Section \ref{spch} (i).

Although a full quantitative inference of formation channels, or a predictive prescription for initial parameters, lies beyond the scope of this work, the mass-shell model can, in principle, be used in conjunction with EOB waveform models or population-synthesis analyses. Such a combined framework could probe how different evolutionary pathways influence e.g. the coalescence energy or the effective angular momentum at coalescence. This is highlighted as a natural direction for yet another future study.

\subsection{On the Cross-Correlation with the Einstein Telescope}

The Einstein Telescope (ET) is a proposed, underground third-generation GW interferometric detector, which intends to improve from its LIGO/VIRGO/KAGRA predecessors instrumentally, constructively, as well as in the analytical effort \cite{ET:2025xjr}. Instrumentally, it is proposed to consist of three LIGO-like detectors, forming an equilateral triangular shape with an arm length of 10 km (compared to e.g. VIRGO's 3-km, L-shaped arm lengths). Upon these three detectors, two interferometers would be constructed: one would measure low-frequency GW signals ($2\sim40$ Hz) and the other high-frequency signals (into the kHz range).  This intertwines with the proposed analytical efforts of the ET collaboration: the waveform modeling of these wide-frequency-range GWs, CB properties across IMR, pre-merger probing of e.g. NS internal structure, etc. We can naturally extend the CB mass-shell model to potential detections made by the ET, more specifically from the energetics perspective via the coalescence energy as Eq. (\ref{erad}) -- or the reduced Kretschmann scalar form -- and via the shell surface energy via Eq. (\ref{energy}).

The mass-shell coalescence energy via Eq. (\ref{erad}) entails a normalized oribital speed component that turns the energy to zero at the upper speed limit of $v_C\leq0.447c$. Given this tangential speed is gauged by the total mass horizon radius and half of the GW's frequency, we can gauge the upper limit of the GW frequency according to the model as
\begin{equation}
f_\mathrm{GW,max}\leq 0.447\frac{c^3}{GM}.
\end{equation}
Provided the total mass of the CB can range within the order of $\sim10^{30}$ kg for BNSs and up to $\sim10^{32}$ kg for BBHs, the ballpark order of magnitude for the maximum GW frequency is within $10^3\sim10^5$ Hz, into the kHz range and below the MHz range. This reinforces the model's versatility for ET-grade high-frequency ranges, as it also opens the door to GW events with coalescence frequencies below 1 MHz.

For low-frequency ranges, supposing them to be coalescence frequencies, the dominating term in Eq. (\ref{erad}) is the residual Newtonian limit $E(t_C)\simeq\pi\alpha\mu c^2/6$. In such a case, the CB in question would behave asymptotically as though under classical Newtonian physics. However, if these low frequencies are produced during the (nearly)-circular pre-merger inspiral phase of CBC, we instead refer to the shell surface energy via Eq. (\ref{energy}), where $\rho=\rho(t)$ has a decreasing rate of change and $a^2/\rho^2=v(t)^2/c^2$ has an increasing rate of change. In essence, we can track the morphology of the CB mass-shell from a ``wide", slow-rotating initial state to a compact, fast-rotating final state, and therewith the time-dependent, pre-merger GW energy radiated. For the latter, one can probe the potential of using the mass-shell waveforms via Eqs. (\ref{hp2}) and (\ref{hc2}) to define the first- and second-order rates of change in $\Omega$ by relating $E(t)\propto h_+^2+h_\times^2$.

\section{Concluding Statements}\label{sect:concl}

In this work, we modeled compact binary coalescence as a rotating, contracting mass shell model. To be in alignment with the very successful effective one-body model, the mass shell peak contraction is the total mass horizon, at the effective moment in CBC where the two CB masses touch surfaces. From the respective quadrupole geometry, the TT-gauged waveforms encode variations in orbital frequency and CB separation while conserving angular momentum. To determine the energy radiating from the mass shell surface as radiating GWs, a variational methodology is employed on the Einstein field equations, assuming the CB mass shell model obeys a Kerr metric Ansatz. Through this methodology, the surface energy at coalescence was obtained to depend on the reduced mass, the symmetric mass ratio, and the normalized rotational speed at merger.

The model calculated anticipated energy values that are close to the Gravitational Wave Transient Catalogue values, with room for improvement resting in the systematic revisions. Thus, the revised model utilizing a reduced Kerr Kretschmann scalar as the effective Ricci scalar, which offers a viable semi-analytical tool to approximate GW energy outputs within agreeable 1:1 ratios for GW150914. Future studies should explore this revised framework in greater depth, particularly in the meticulous comparison between anticipated energy values and observed events (see e.g. Ref. \cite{MacKay:2025uyg}). It is also of interest to investigate PN corrections in the CB mass-shell model in the Laplace-Beltrami formalism, to include e.g. tidal and eccentric corrections, and how they, in turn, alter the waveform profiles and wave envelopes via Eqs. (\ref{hp2}) and (\ref{hc2}).

\section*{Acknowledgments}

I thank the referees on their insightful comments, which have further enhanced this work.

\section*{Statement Declarations}

\subsection*{Conflict of Interest}
The author declares no conflicts of interest.

\subsection*{Data Access Statement}
As a theoretical study, this work generates no original data. Data from cited LIGO observations are publicly available.

\subsection*{Ethics Statement}
No ethical issues arise, as no test subjects are involved. This paper adheres to academic integrity.

\subsection*{Funding Statement}
This work received no funding.

\appendix

\section{Bianchi Identity Sanity Check}\label{ap1}

Given $G_{\mu\nu}\propto T_{\mu\nu}$ via Eq. (\ref{tens1}), standard GR enforces the vanishing covariant derivative on both sides of the EFEs, upholding the energy-momentum conservation $\nabla^\mu T_{\mu\nu}=0$ and the Bianchi identities $\nabla^\mu G_{\mu\nu}=0$ concurrently. Through $\nabla^\mu G_{\mu\nu}=0$, one has the reduced Bianchi identity of the form
\begin{equation}
\nabla^\mu R_{\mu\nu}=\frac{1}{2}\nabla_\nu R,
\end{equation} 
with metric compatibility $\nabla^\mu g_{\mu\nu}=0$ readily implied. This relation holds for all metrics that satisfy the EFEs, including the Kerr metric adopted in the variational approach.

The present work utilizes the Christoffel-symbol-free, coordinate-dependent Laplace-Beltrami formulation of the Ricci tensor, which is provided generally (i.e., independent of one's choice of metric and coordinates) as Eq. (\ref{ricci}) with the lower order terms not explicitly defined. These lower order terms are respectively a first-order covariant derivative of a rank-1 vector $V_\nu$ and a zeroth-order auxiliary rank-2 tensor $A_{\mu\nu}$. The Ricci tensor can therefore be written as
\begin{equation}
R_{\mu\nu}=-\frac{1}{2}\Delta^\mathrm{LB} g_{\mu\nu}+\nabla_{(\mu}V_{\nu)}+A_{\mu\nu},
\end{equation} 
where $\nabla_{(\mu}V_{\nu)}=\nabla_\mu V_\nu+\nabla_\nu V_\mu$ is a symmetric combination. The corresponding Ricci scalar is
\begin{equation}
R=-\frac{1}{2}g^{\mu\nu}\Delta^\mathrm{LB} g_{\mu\nu}+2\nabla^\mu V_\mu +A,
\end{equation}
implying $g^{\mu\nu}\nabla_\mu=\nabla^\nu$ and $g^{\mu\nu}A_{\mu\nu}=A$. This form is manifestly coordinate-independent.  Applying a covariant derivative to the Laplace-Beltrami representation of $R_{\mu\nu}$ gives
\begin{equation}\label{bian}
\nabla^\mu R_{\mu\nu}=-\frac{1}{2}\nabla^\mu \Delta^\mathrm{LB} g_{\mu\nu}+\nabla^\mu\nabla_{(\mu}V_{\nu)}+\nabla^\mu A_{\mu\nu}.
\end{equation}

We now verify that the reduced Bianchi identity remains satisfied within this formalism, confirming consistency without imposing additional gauge constraints. It should be noted that in the main text, the Laplace-Beltrami operator acts explicitly on the Kerr metric components in a coordinate-dependent form, provided as Eq. (\ref{lbk}) and therefore providing non-zero measurables. Here, it is treated in its coordinate-independent form $\Delta^\mathrm{LB}=\nabla_\alpha\nabla^\alpha$ to verify general covariance and metric compatibility. By this caveat, the leading second-order term vanishes. Therefore, only the lower-order terms remain in Eq. (\ref{bian}). For the Ricci scalar, its covariant derivative in the Laplace-Beltrami formalism reads
\begin{equation}
\nabla_\nu R=-\frac{1}{2}\nabla_\nu \left(g^{\mu\nu}\Delta^\mathrm{LB} g_{\mu\nu}\right)+2\nabla_\nu\nabla^\mu V_\mu +\nabla_\nu A.
\end{equation}
The leading order term is readily zero due to metric compatibility. Equating both sides of the Bianchi identity therefore gives the following:
\begin{eqnarray}
&\nabla^\mu\nabla_{(\mu}V_{\nu)}+\nabla^\mu A_{\mu\nu}=\frac{1}{2}\left(2\nabla_\nu\nabla^\mu V_\mu +\nabla_\nu A \right)\\\nonumber
&\Rightarrow\quad \nabla^\mu\nabla_{(\mu}V_{\nu)}=\nabla_\nu\nabla^\mu V_\mu\quad\mathrm{and}\quad\nabla^\mu A_{\mu\nu}=\frac{1}{2}\nabla_\nu A.
\end{eqnarray}
One can see readily that the Bianchi identity is satisfied for the zeroth-order auxiliary tensor $A_{\mu\nu}$. 

For the first-order term, one can utilize the symmetric combination to yield an emergent Laplace-Beltrami operator acting on the 4-vector:
\begin{equation}
\nabla^\mu \nabla_\mu V_\nu:=\Delta^\mathrm{LB}V_\nu=\nabla_\nu\nabla^\mu V_\mu-\nabla^\mu\nabla_\nu V_\mu.
\end{equation} 
The right-hand side can be expressed as a commutator of covariant derivatives acting on the 4-vector $V_\mu$:
\begin{equation}
\nabla_\nu\nabla^\mu V_\mu-\nabla^\mu\nabla_\nu V_\mu=g^{\mu\rho}[\nabla_\nu,\nabla_\rho]V_\mu=g^{\mu\rho}R_{\sigma\mu\nu\rho}V^\sigma,
\end{equation}
where the rightmost expression follows from the definition of the Riemann tensor. As $g^{\mu\rho}R_{\sigma\mu\nu\rho}$ defines the Ricci tensor $R_{\sigma\nu}$, the first-order Bianchi identity in the Laplace-Beltrami formalism gives a curvature-coupled, inhomogeneous wave-like equation (a form of the Weitzenb\"ock identity) for the 4-vector:
\begin{equation}
\Delta^\mathrm{LB}V_\nu=R_{\sigma\nu}V^\sigma.
\end{equation}

The derivation above confirms that the Laplace-Beltrami representation of the Ricci tensor satisfies the contracted Bianchi identity, which is only possible due to the lower-order terms.  The first-order Bianchi identity, if one views this from a physical perspective, coincides in form with the covariant wave equation for a vector field in curved spacetime. If $V_\nu$ were regarded as an independent dynamical quantity rather than an auxiliary term in the Laplace-Beltrami expansion, this relation would describe the propagation of a curvature-coupled, spin-1-like field. In this sense, the Laplace-Beltrami decomposition of the Ricci tensor hints at a natural geometric origin for an emergent rank-1 field associated with curvature -- one that is ordinary overshadowed by the leading, Laplacian-type operation on the metric tensor itself. 




\end{document}